\documentclass[fleqn,usenatbib]{mnras}

\usepackage[T1]{fontenc}
\usepackage{ae,aecompl}

\usepackage{graphicx}	

\newcommand{\hii}{H$\,${\sc ii}\ }
\newcommand{\changed}[1]{#1}  

\title[NED and SIMBAD in nearby galaxies]{Comparing NED and SIMBAD classifications across the contents of nearby galaxies}

\author[Kuhn, Shubat, \& Barmby]{
L. Kuhn$^{1,2}$, 
M. Shubat$^{1,3}$
P. Barmby$^{1,4}$
\thanks{E-mail: pbarmby@uwo.ca}
\\
$^{1}$Department of Physics \& Astronomy, Western University, London, Canada N6A 3K7\\
$^{2}$Department of Physics \& Astronomy, The University of British Columbia, Vancouver, Canada V6T 1Z1\\
$^{3}$Department of Computer Science, Western University, London, Canada N6A 3K7\\
$^{4}$Institute for Earth and Space Exploration, Western University, London, Canada N6A 3K7\\
}

\date{Accepted XXX. Received YYY; in original form ZZZ}

\pubyear{2022}

\DeclareRobustCommand{\VAN}[3]{#2}
\let\VANthebibliography\thebibliography
\def\thebibliography{\DeclareRobustCommand{\VAN}[3]{##3}\VANthebibliography}

\begin{document}
\label{firstpage}
\pagerange{\pageref{firstpage}--\pageref{lastpage}}
\maketitle

\begin{abstract} 

Cataloguing and classifying celestial objects is one of the fundamental activities of observational astrophysics.
In this work, we compare the contents of two comprehensive databases, the NASA Extragalactic Database (NED) and Set of Identifications, Measurements and Bibliography for Astronomical Data (SIMBAD) in the vicinity of nearby galaxies.
These two databases employ different classification schemes -- one flat and one hierarchical -- and our goal was to determine the compatibility of classifications for objects in common.
Searching both databases for objects within the respective isophotal radius of each of the $\sim 1300$ individual galaxies in the Local Volume Galaxy sample, 
we found that on average, NED contains about ten times as many entries as SIMBAD and about \changed{two thirds} of SIMBAD objects are matched by position to a NED object, at 5\arcsec\ tolerance.
These quantities do not depend strongly on the properties of the parent galaxies.
We developed an algorithm to compare individual object classifications between the two databases and found that \changed{88\%} of the classifications agree; we conclude
that NED and SIMBAD contain consistent information for sources in common in the vicinity of nearby galaxies.
Because many galaxies have numerous sources contained only in one of NED or SIMBAD, researchers seeking the most complete picture of an individual galaxy's contents
are best served by using both databases.
\end{abstract}


\section{Introduction}

For thousands of years, astronomers have been attempting to catalogue, classify, and understand the contents of the universe as observable from the Earth.
Technological advances have allowed us to see deeper, farther, and over broader regions of the electromagnetic spectrum.
Combining observations made with disparate facilities and across epochs has been and continues to be a central challenge of the field.
While comprehensive all-sky surveys produce uniform information, many astronomical investigations focus on particular objects or classes of objects, and there is a need to combine the results of multiple surveys with previous observations to gain a more complete understanding.

Modern astrophysics has two major databases which attempt to compile astrophysical measurements of the properties of individual celestial objects, as published in the professional literature: the NASA Extragalactic Database \citep[NED; ][]{mazzarella_evolution_2017} and Set of Identifications, Measurements and Bibliography for Astronomical Data \citep[SIMBAD; ][]{wenger_simbad_2000}.
NED was established in 1988 \citep{helou1988} with the goals of collecting information from the published astronomical literature on objects outside the Milky Way.
These include summaries of published data for named objects, searches for objects by sky position, and database-style search-and-selection for samples of extragalactic objects.
Since its inception, NED has evolved from ``an object reference database  into a data-mining discovery engine'' \citep{mazzarella_evolution_2017}.
Recent enhancements to NED have included cross-identifying objects between large sky surveys such as AllWISE and SDSS \citep{ogle2015}  and cataloguing redshift-independent distances to nearby galaxies \citep{steer2017}.
As of \changed{April 2022 (data release 32.31),
NED contains $1.11\times10^9$ distinct objects, $1.52\times10^9$ multiwavelength associations, $1.40\times10^{10}$ photometric data points, and $5.55\times10^7$} links between an object and literature reference.%
\footnote{\url{http://ned.ipac.caltech.edu/CurrentHoldings}}
It contains $5.02\times10^5$ detailed classifications for $2.30\times10^5$ distinct objects.

SIMBAD was established in the early 1980s as a merger of two stellar catalogues \citep{wenger_simbad_2000} and extended to include galaxies in 1983 \citep{egret1983}. 
It specifically excludes the Sun and solar system bodies \citep{wenger_simbad_2000} but otherwise attempts to provide a comprehensive compilation of the literature.
As of \changed{June 2022 (data release 4 1.8), SIMBAD contains $1.33\times10^7$ distinct objects, $5.30\times10^7$ identifiers, and $3.02\times10^7$} links between an object and literature reference.%
\footnote{\url{https://simbad.u-strasbg.fr/simbad/}}

With NED originally designed for whole-galaxy properties and SIMBAD for individual star properties, their overlap in the nearby-galaxy distance regime provides a particularly interesting realm for exploration and comparison.
Scientifically, nearby galaxies provide an intermediate regime between the detailed study of (the contents of) the Milky Way and the study of integrated galaxy properties at higher redshifts.
Neither NED nor SIMBAD was originally designed to deal with the multi-scale systems that are nearby galaxies.
Unlike individual stars or distant galaxies, nearby galaxies can be resolved into both individual objects, such as stars and molecular clouds and objects which are themselves further resolvable, such as star clusters. 
Nearby galaxies show us the results of galaxy evolution in detail, and understanding their contents is important in relating the properties of galaxies on the micro- and macro-scales.

One critical aspect of understanding the contents of nearby galaxies is classifying objects that belong to them.
While classification in itself does not necessarily lead to detailed physical understanding, grouping related phenomena -- from variable star light curves to galaxy morphologies to quasar spectra -- can yield information about the number and extent of physical processes driving the properties of objects.
NED and SIMBAD use quite different classification schemes.
Comparing the two schemes shows that NED's flat classification scheme is somewhat more detailed for extragalactic objects (galaxy and active galaxy types) while SIMBAD's hierarchical scheme is much more detailed for stars and stellar systems; this is understandable in light of the two databases' original goals.

A researcher seeking to understand the contents of nearby galaxies might well make use of both NED and SIMBAD, or might assume that only one is needed because their contents are identical.
To our knowledge, no published work directly compares the two databases, so the assumption above has not been tested.
In this work we compare the contents of NED and SIMBAD in the vicinity of nearby galaxies, an area where the two databases overlap but for which neither was originally designed.
As a comparison sample we use the Local Volume Galaxy sample defined by \citet{kara13}.
Our goal is to understand how much of the nearby-galaxy content of NED and SIMBAD is in common, whether and how their classifications of individual galaxy constituents differ, and how these differences vary with galaxy properties.
While the data science literature discusses algorithms for comparing two flat classification schemes, or two hierarchical schemes \citep[e.g.][]{brucker2011,ulanov2011, hutchison2012}, comparing a flat scheme to a hierarchical scheme has received little attention and is part of the novelty of this work.
Our overall intent is to provide a guide to other researchers seeking to use NED and SIMBAD in the study of nearby galaxies.

\section{Methods}

We created a Python package, ``galaxy data mines'' (gdmines), to perform NED/SIMBAD comparisons. 
The package makes use of the SciPy infrastructure, including Matplotlib \citep{matplotlib2007},
NumPy \citep{numpy_2011} and Pandas \citep{pandas2010}. 
The astronomy library Astropy \citep{astropy2018} was utilized for table processing and its affiliated package Astroquery \citep{astroquery2018} was employed to query both NED and SIMBAD. 
Using Astropy and Astroquery, the gdmines tool queries NED and SIMBAD dynamically for objects within a user-specifiable radius of any sky coordinate, or astronomical object resolvable by Astropy to a sky coordinate.

\subsection{Matching NED and SIMBAD sources}
\label{sect:ns_match}

NED and SIMBAD often draw from the same literature sources; however they use slightly different nomenclature conventions.
For example, NED and SIMBAD each contain a source in NGC~6822 at nearly the same coordinates: NED lists a Wolf-Rayet star called ``NGC 6822:[AM85] 11'' while
SIMBAD lists an emission-line star ``[AM85] NGC 6822 11''.
To an astronomer it's clear that both names refer to object \#11 in the list of \citet{am85}, but the variety in naming conventions used by both databases and original authors makes it difficult to develop an algorithm to match objects within galaxies by name.
Thus our matching procedure relies only on sky coordinates. 
It also uses a single distance tolerance for matching objects between the two databases. 
While different individual objects and/or catalogues may have different positional uncertainties, the literature sources from which NED and SIMBAD are derived do not consistently report such uncertainties, so using a single tolerance for matching is more practical.

Depending on the application, astronomical source lists can be cross-matched by position in a number of different ways: one-to-one, one-to-several, probabilistic, etc. \citep{Wilson_2017,Budavari_2008}.
In gdmines a symmetric match is used: for each object in both of the input lists, the closest object in the other list (within a tolerance) is found.
Objects in different lists are considered a match only if each of them are the other's closest match. 
While this approach could potentially miss some true matches, as instances of slight positional offsets of counterpart objects between both source lists (as detailed in \autoref{sec:galaxy_class}) could allow for an object to be positionally cross-matched to an incorrect analogue, it ensures that a given object is only counted once in a list of matches. 
We choose to prioritise preventing repeated counts of a single object as it could lead to overestimating the total number of calculated matches as well as an uncertainty in the number of true matches enumerated.
 
Once objects are retrieved from NED and SIMBAD, the two object lists are matched using a user-specified matching tolerance.
No filtering on object type or other properties is performed before the matching takes place.
This means that objects physically unrelated to the target galaxy (e.g. foreground Milky Way stars or background galaxies) are retained. 
This is by design: such objects are potential contaminants and their correct classification is therefore important.
After matching, a quantitative summary of the process is computed, giving the number and fraction of matched and unmatched objects within each of the input lists.
The classifications of the matched objects are then compared as detailed below.

\subsection{Comparing classifications}
\label{sec:compare_class}

\begin{figure}
\includegraphics[width=\columnwidth]{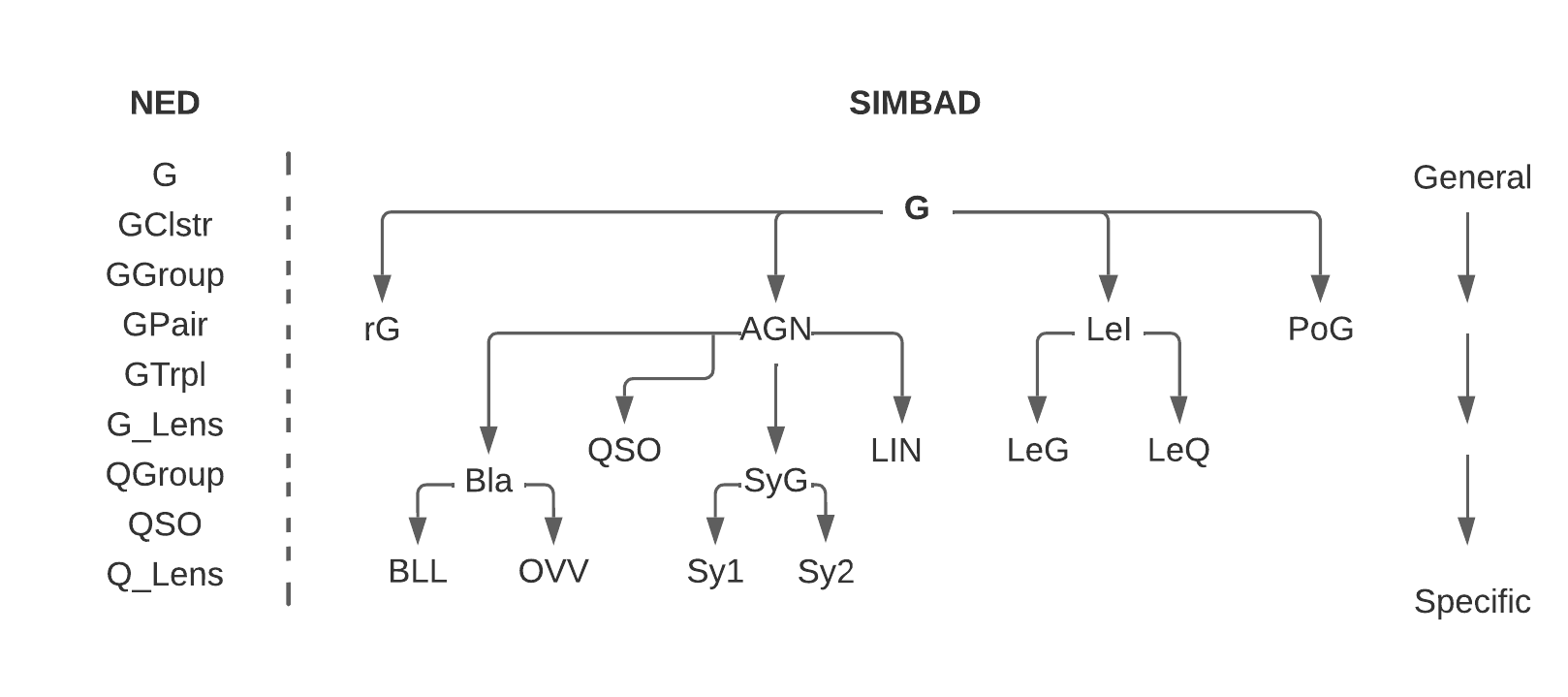}
\caption{
\changed{
Comparison of excerpts from the NED (left) and SIMBAD (right) classification schemes. The SIMBAD scheme has multiple levels, while the NED scheme is a single-level list of categories.
Refer to NED and SIMBAD object classification lists
for extended explanations of all object abbreviations.
Note that this figure refers to a previous version of the SIMBAD hierarchy; see Appendix A for details.
}}

\label{fig:simbad_class}
\end{figure}

The NED classification scheme consists of 63 different categories, listed in \autoref{tab:ned_classes}. 
The NED categories are separated into different types: classified extragalactic (galaxies and structures containing galaxies from pairs to clusters, QSOs, gravitational lenses, absorption and emission line systems), components of galaxies (e.g. supernovae, star clusters, carbon stars, emission nebulae), and unclassified objects identified only by their detection wavelength (e.g. radio sources, infrared sources, etc).
The SIMBAD classification scheme (\autoref{fig:simbad_class}), in contrast to NED's, is hierarchical---classifications include categories and sub-categories, for a total of 167 final categories in a scheme that is four levels deep. 
Further details on the SIMBAD scheme are given by \citet{oberto2018}.

\begin{figure*}
\includegraphics[width=\textwidth]{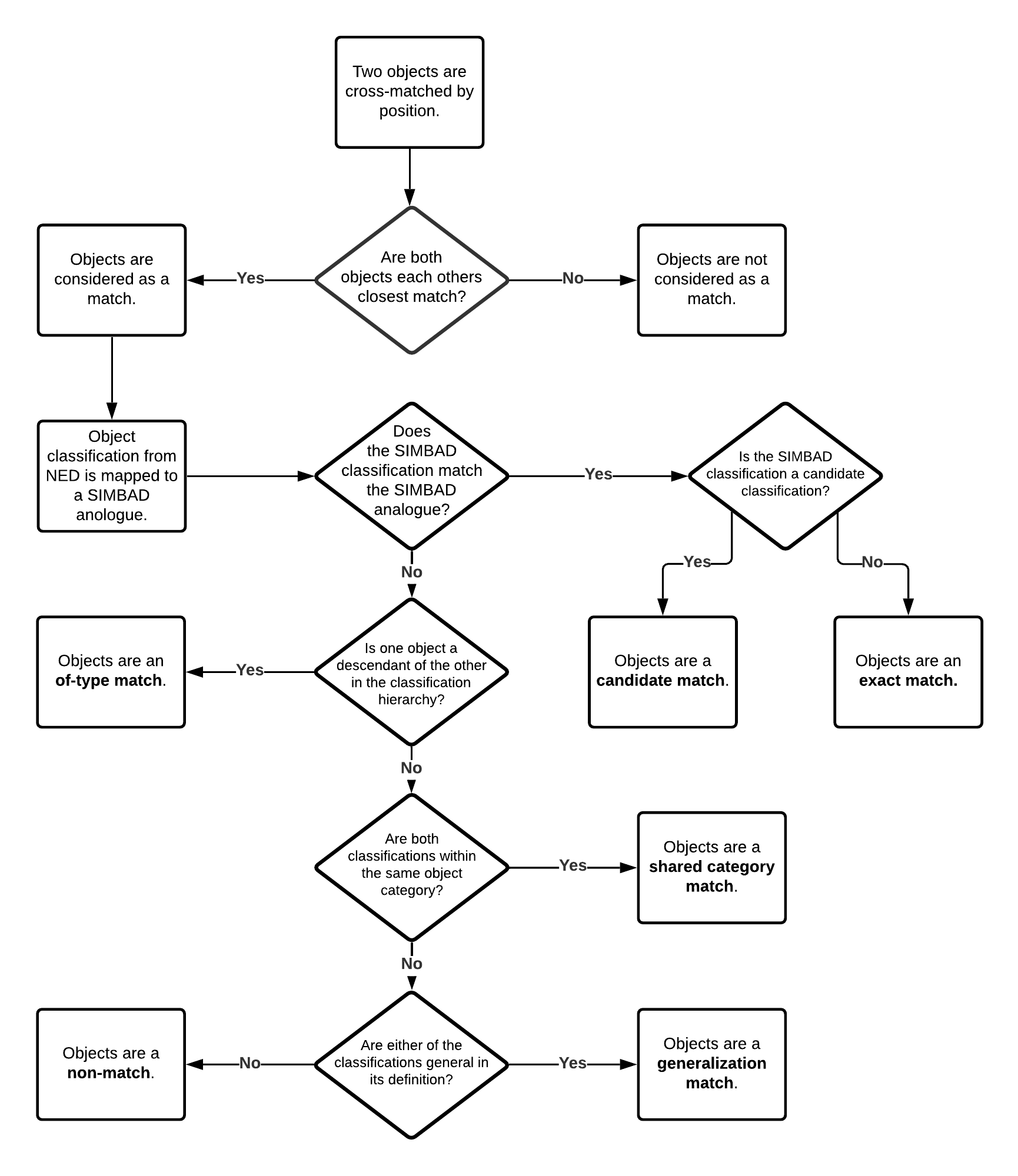}
\caption{NED/SIMBAD comparison algorithm}
\label{fig:ns_compare_alg}
\end{figure*}

SIMBAD's hierarchical classification scheme is effectively a superset of the flat scheme employed by NED. 
Our algorithm defines a mapping between the two schemes in order to compare the classifications for individual objects, assigning them as being either strong matches, weak matches or non-matches. 
\autoref{fig:ns_compare_alg} summarizes the comparison algorithm; the match categories are described in more detail below.

Strong classification matches are those where we consider the classifications to be identical or basically identical.
Strong matches are divided into three distinct sub-classes: exact matches, candidate matches, and of-type matches.
{\em Exact matches} occur when NED and SIMBAD both give the same class to an object. 
Syntactic differences between corresponding classes, for example, ``star cluster'' (*Cl) in NED and ``cluster of stars'' (Cl*) in SIMBAD, are ignored.
{\em Candidate matches} occur when the NED classification matches with a SIMBAD `candidate classification', for example a NED
``cluster of galaxies'' (GClstr) and SIMBAD ``possible cluster of galaxies'' (C?G).
NED's classification scheme does not include candidate objects.
{\em Of-type matches} are matches where one object is a descendant of the other in the classification hierarchy, for example ``star cluster'' (*Cl) in NED and ``open cluster'' (OpC) in SIMBAD. 

Weak  classification matches are those which are related, but not as closely as for strong matches. 
These are less confident match types, but still compatible.
Weak matches are divided into two distinct sub-classes: shared matches and generalization matches.
{\em Shared category matches} occur when the two classifications share a common ancestor in the SIMBAD classification hierarchy, and therefore are in the same object category.
An example is an object classified by NED as a ``nebula'' (Neb, which is mapped to the SIMBAD analogue ``cloud'' Cld), and by SIMBAD as a ``supernova remnant candidate'' (SNR?), both of which share the same object category of interstellar matter. 
{\em Generalization matches} occur when one of NED or SIMBAD gives an object a very general definition, which is not necessarily incompatible with other classifications but is also not definitive.
An example is a NED classification as ``X-ray source'' (XrayS) and a SIMBAD classification as a galaxy (G).

The final category of classification matches is non-match, the label given to overlapping objects for which none of the above conditions apply.
An example is a NED classification as an ``star'' (*) and a SIMBAD classification as an ``galaxy'' (G).
It is important to note that a ``non-match'' class given to a pair of objects does not guarantee that the given classifications disagree, it only indicates that currently there is no defined match type for the given relationship.

\subsection{Galaxy sample}
\label{sec:galaxy_class}

We compared the contents of NED and SIMBAD for the sample of nearby galaxies described by \citet{kara13} --- the Local Volume Galaxy catalog, or LVG hereafter.
This catalogue has been updated since the above publication; the edition dated 2020 August 12 was used, downloaded from \url{http://www.sao.ru/lv/lvgdb/}.
The catalogue contains 1246 objects, located within 11~Mpc of the Milky Way or measured to have a corrected radial velocity $v_{\rm LG}< 600$~km~s$^{-1}$ with respect to the centroid of the Local Group. 
A detailed description of the LVG and its construction appears in \citet{kara13}.
The majority of galaxies within the volume are classified as spheroidal dwarfs with {\changed luminosities $\sim 10^{-4}L_{\rm MW}$,} and the catalogue includes over a dozen groups, similar in size and population to the Local Group.
Other typical properties of objects catalogued within the LVG database include \changed{an average radius of 2.2 kpc, an average mass of $10^{8.1}$~M$_{\odot}$ within the Holmberg radius}, and the most common galaxy morphology type corresponding to armless Magellanic irregulars.  
From the LVG, we remove the entries for the Milky Way, Bootes III, M31, M33, Antlia 2, Crater 2, the Sagittarius dSph, and the Large and Small Magellanic Clouds.
The large angular extents of these objects means that they are more likely to be contaminated by foreground sources and also less typical of nearby galaxies in general. 
Studying the NED and SIMBAD contents of these galaxies separately would be a possible follow-up to the present work.
Additionally, we remove 11 entries from the LVG which did not have a measurement of major linear diameter. The removals result in a final galaxy sample size of 1227 objects.

\begin{figure}

\includegraphics[width=\columnwidth]{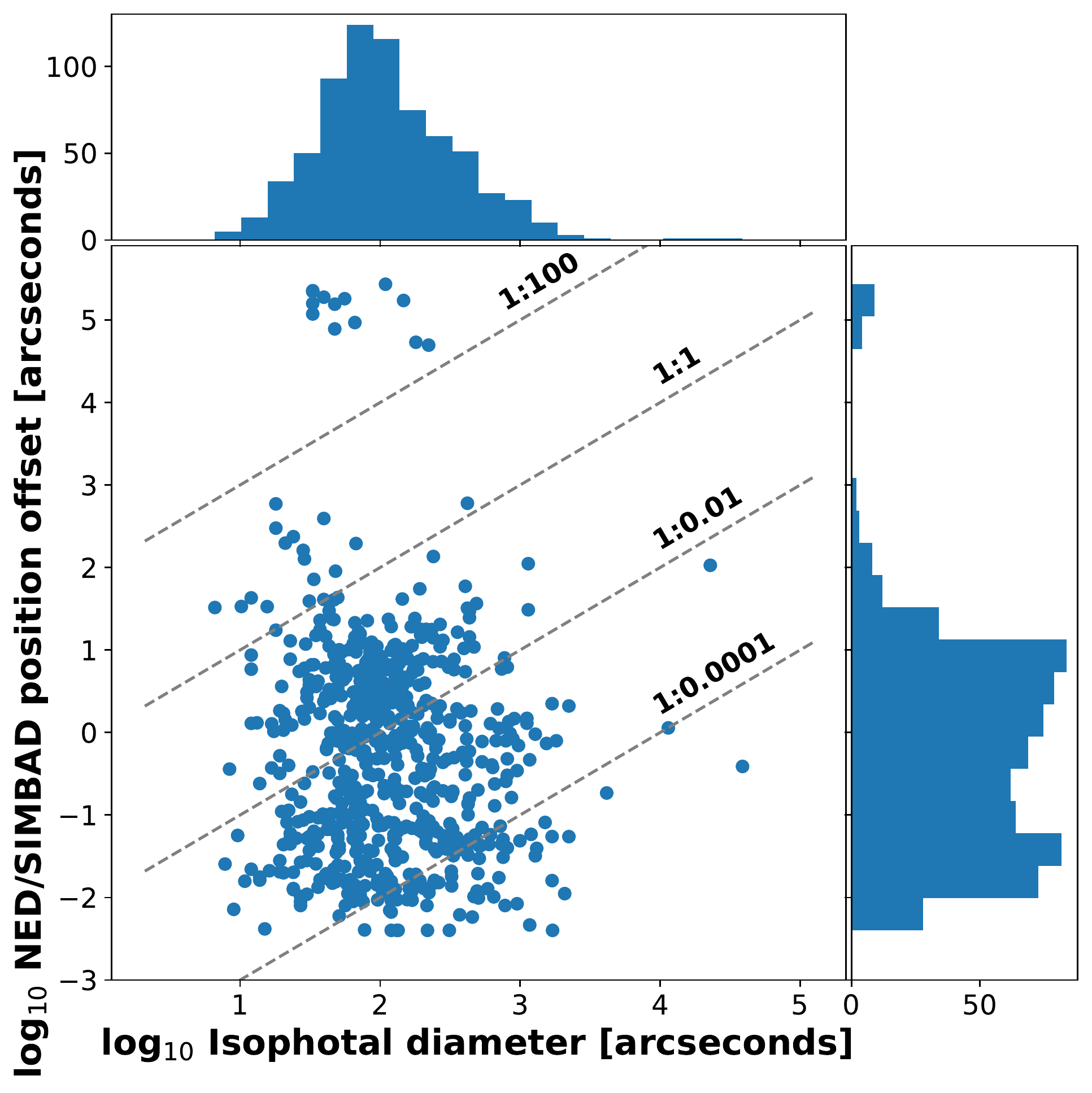}
\caption{Position offsets between NED and SIMBAD as a function of isophotal diameter for 698 nearby galaxies queried by name in both databases. 
Five galaxies with position offsets of 0 are omitted (KDG~058, KDG~065, UMa~I, Aquarius~2, Tucana~III).
The histograms coupled with the horizontal and vertical axes represent the number of galaxies with a given position offset and isophotal diameter respectively. \changed{The dashed lines indicate the ratio of the position offset to the isophotal diameter of the galaxy.}} 
\label{fig:ns_pos_diff}
\end{figure}

A first step in comparing the contents  NED and SIMBAD for nearby galaxies is to compare the entries for the galaxies themselves.
As with many other areas of astronomy, the nomenclature of nearby galaxies is complex and inconsistent.
For nearby galaxies with a large angular extent, there is an additional complication that even their positions on the sky may not be well-defined.
Our initial experiments showed that, of the 1235 LVG galaxies with a recorded measurement of major linear diameter, 698 were listed in SIMBAD, and 864 were listed in NED, by the LVG name.
Even where the same galaxy is listed in both databases, it may be at a different position: for example the \changed{(epoch J2000)} central position of the dwarf irregular galaxy IC~10 is given as $00^{\rm h}20^{\rm m}17.34^{\rm s}, +59^{\circ}18^{\prime}13.6^{\prime\prime}$ in NED and $00^{\rm h}20^{\rm m}23.16^{\rm s}, +59^{\circ}17^{\prime}34.7^{\prime\prime}$ in SIMBAD. 
This distance of nearly an arcminute between these positions is significant for a galaxy whose isophotal diameter is 13\farcm5. 
\autoref{fig:ns_pos_diff} shows the distribution of position differences between NED and SIMBAD for LVG galaxies searched by name in the two databases. 
There is no correlation between the size of the positional offset and the galaxy size, and
the median offset between positions in the two databases is quite small (0\farcs68). 
There are 13 objects with position offsets $>1000$~arcsec;  in these cases it appears that the NED and SIMBAD name resolvers are associating the names with different objects, for example
``Virgo I'' is a dwarf galaxy according to NED but the Virgo galaxy cluster according to SIMBAD. 11 of these 13 contain names with the prefix `[KK2000]'; 
NED has the same coordinates as in the LVG for these objects.

The result of the name and positional comparison is that we chose to query NED and SIMBAD by galaxy position as listed in the LVG catalog, and not by galaxy name.
For each galaxy, the object query radius was set to $a_{26}$, the angular diameter of the $B=26.5$ isophote listed by the LVG.
Although this query radius could exclude some outlying members of a galaxy, such as distant globular clusters, it provides a size measurement that is consistently-determined between galaxies.

\changed{A consistent tolerance was also used when matching objects between NED and SIMBAD.
Offsets between different measurements of the same object are likely to be dominated by astrometric accuracy and localization precision, not galaxy-distance-dependent physical distances, so using a constant tolerance value is justifiable.
To determine the tolerance value, we selected several galaxies with large numbers of both NED and SIMBAD objects (M101, NGC~6822, NGC~5128) and examined the change in the number of matches as the tolerance was increased.
The number of matches increased rapidly with tolerance at tolerance values below 2~arcsec, and appeared to saturate at tolerance values above 5~arcsec.
We thus chose to use 5~arcsec as the consistent value across galaxies.
Although astrometric accuracy may in many case be better than 5~arcsec, the number of spurious matches should be reduced by the symmetric matching procedure used by gdmines (see \autoref{sect:ns_match}).}

\section{Results}

\subsection{NED versus SIMBAD: objects per galaxy}

For the galaxies in the LVG sample, NED and SIMBAD contain a total of \changed{1 165 663 and 112 287} objects within the $a_{26}$ radius, \changed{a ratio of $N_{\rm N}/N_{\rm S} = 10.4$}.
Most of these objects are associated with only a few galaxies: 50\% of the NED objects are associated with the 22 most populous galaxies, and 50\% of the SIMBAD objects are associated with the \changed{17} most populous galaxies.
While a handful of galaxies do contain more SIMBAD than NED objects, including several of the Andromeda satellites and a few distant members of the Local Group, NED otherwise contains a significantly greater number of objects per galaxy.
Averaged on a per-galaxy basis, \changed{the ratio $N_{\rm N}/N_{\rm S} = 26.8\pm 1.2$} where the uncertainty is the standard error of the mean.

\begin{figure*}
\includegraphics[width=0.9\textwidth]{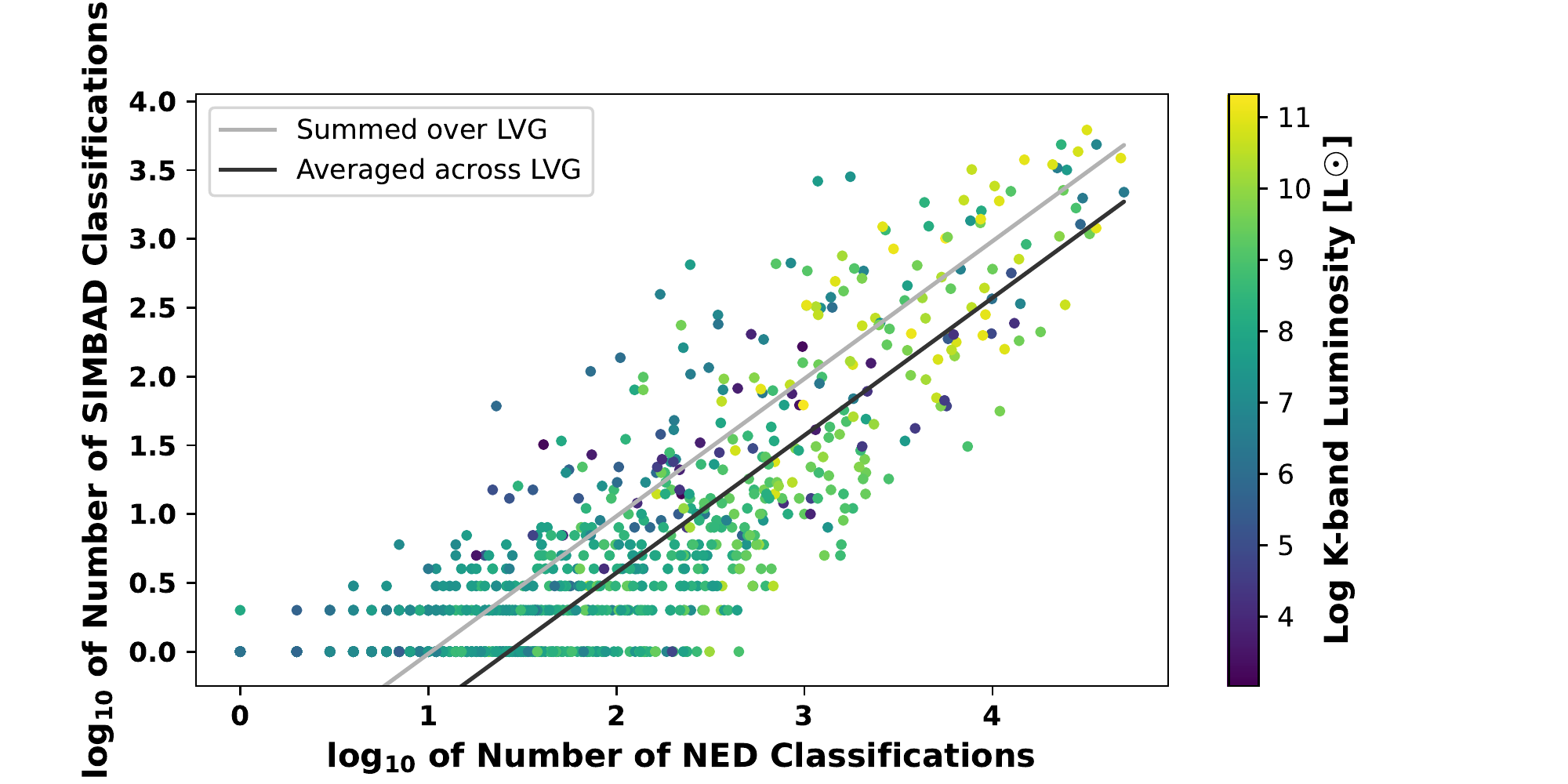}
\caption{Comparison of the number of NED and SIMBAD objects for nearby galaxies, as a function of the galaxy $K_{\rm S}$-band luminosity in solar units, corrected for extinction. The solid grey line ($\log_{\rm 10} N_S = \log_{\rm 10} N_N - 1.02$) and the solid black line ($\log_{\rm 10} N_S = \log_{\rm 10} N_N - 1.43$) illustrate the ratio of NED/SIMBAD objects determined by summing over and averaging across the galaxy sample respectively. The \changed{34 galaxies} with no objects in one or both of NED or SIMBAD are not represented in the figure.}
\label{fig:ns_compare_number}
\end{figure*}

\autoref{fig:ns_compare_number} compares the number of NED and SIMBAD objects returned from the queries for individual nearby galaxy positions as a function of the galaxy's $K_{\rm S}$-band luminosity listed in the LVG.
There is a clear trend between the number of NED and SIMBAD objects returned for each queried galaxy, as might be expected. 
As the figure shows, the per-galaxy average is driven by the many small galaxies
with few NED and SIMBAD objects, while the sum is driven by the small number of galaxies with many more NED and SIMBAD objects.
The colour bar to the right of the figure describes the $K_{\rm S}$-band luminosity of each queried galaxy. The most luminous galaxies (i.e. the dots yellow and light green in colour) collect in the top right portion of the plot, indicating that more luminous galaxies tend to also contain a greater number of NED and SIMBAD objects. This is expected, as galaxies with higher luminosities are also generally larger and are studied more intensively, thus resulting in both a larger number of objects to be identified per galaxy as well as a greater possibility that these objects will be included in the professional literature referenced by NED and SIMBAD.

\begin{figure*}
    \includegraphics[width=0.9\textwidth]{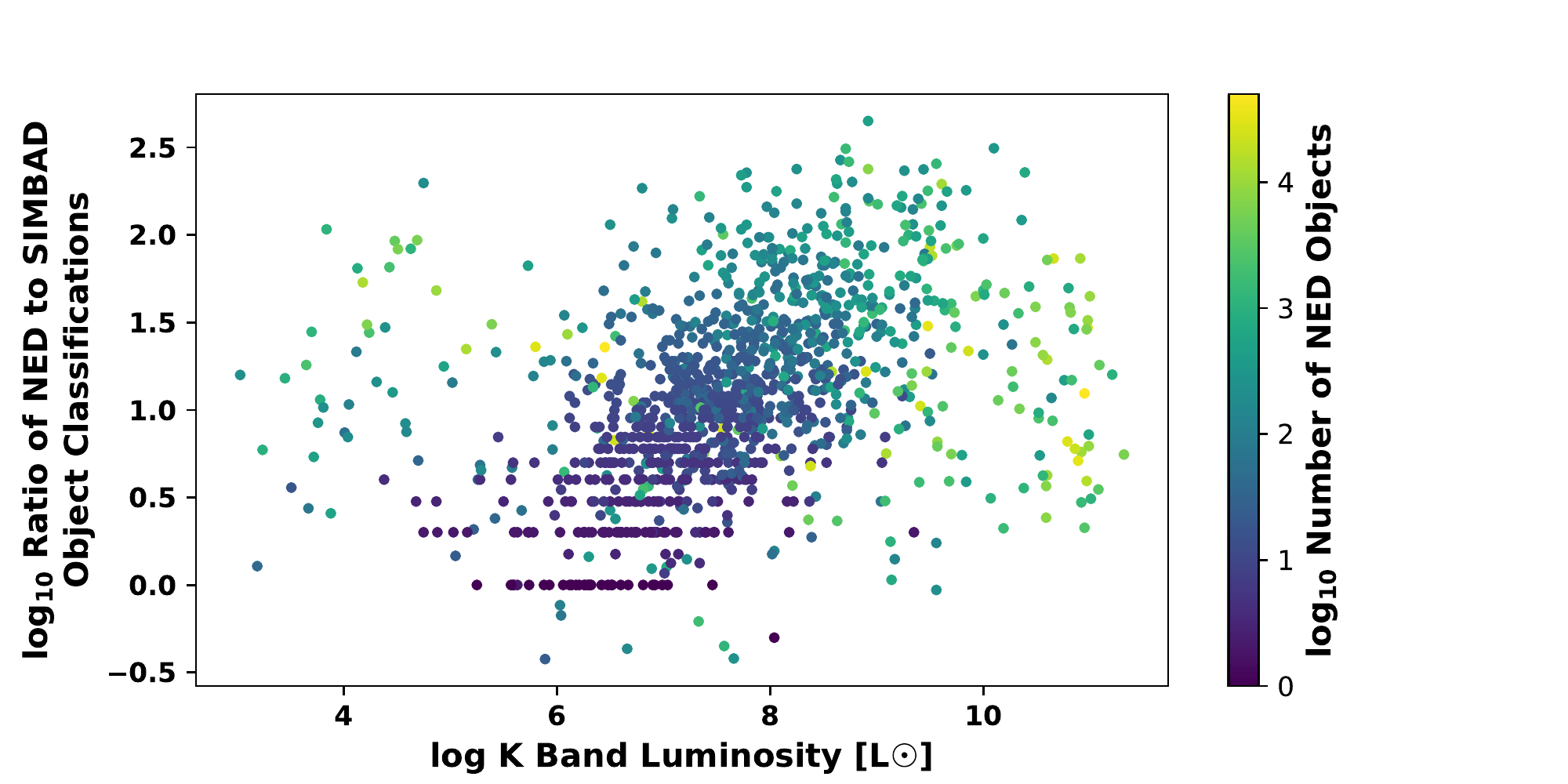}
    \caption{Ratio of the number of NED to SIMBAD objects for nearby galaxies, as a function of the galaxy $K_{\rm S}$-band luminosity in solar units, corrected for extinction. The colour bar represents the number of NED objects per galaxy.}
    \label{fig:ns_compare_overlap}
\end{figure*}

\autoref{fig:ns_compare_overlap} illustrates the ratio of the number of NED to SIMBAD objects returned from individual queries of nearby galaxy positions as a function of the galaxy's $K_{\rm S}$-band luminosity listed in the LVG. 
There is a rough correlation between the ratio of NED to SIMBAD object classifications per galaxy and the galaxy's luminosity. 
The colour bar to the right of the figure depicts the number of NED objects classified in each galaxy. 
The galaxies with more NED classifications (represented by yellow and light green dots) tend to ignore the general trend noted above, but rather collect towards the far left and right of the figure. 
This deviation can be understood by considering the nature of the offset galaxies. 
The less luminous galaxies on the far left of the figure represent better-studied very nearby galaxies for which we anticipate that more individual stellar objects are classified, and the more luminous galaxies on the far right of the figure represent the larger galaxies that contain larger populations of some distinct object classes, such as star clusters or supernova remnants.

\changed{For 30} of the input positions, SIMBAD and NED searches both return exactly one object.
Examination of a random sample of these queries shows that this is usually the galaxy itself. 
Because of the different nomenclature and positions used by NED and SIMBAD, automatically removing these ``matches'' from the comparison was not possible. 
Since they are far outnumbered by the galaxy constituents in the overall sample, we do not expect them to substantially bias the comparison results.

For each galaxy, NED and SIMBAD objects were positionally matched according to the scheme described in  \autoref{sect:ns_match}.
Summed across the entire sample of galaxies, \changed{$\sim 6$\% of NED objects were matched to a SIMBAD object and $\sim 67$\% of SIMBAD objects were matched to a NED object. 
Averaged on a per-galaxy basis, these match fractions become $\sim 12$\% and $\sim 84$\%, respectively.}
These match fractions do not depend strongly on either the number of NED objects in a galaxy, the distance of the galaxy, or the sky position, validating our decision to use a single matching radius for all galaxies. 
Additionally, the match fractions have no direct correlation with any intrinsic galaxy properties including luminosity, mass, diameter, or morphology.
The LVG galaxies with large numbers of SIMBAD objects unmatched in NED tend to be well-studied galaxies (e.g. NGC~147, Messier 81, Fornax dSph) where catalogues of individual stars have been incorporated into SIMBAD but not NED.
The LVG galaxies with the largest numbers of NED objects unmatched in SIMBAD tend to be spiral galaxies viewed close to edge-on (e.g. NGC~4236, NGC~4945).
In these cases the NED sources are dominated by large infrared point-source catalogues such as those from the 2MASS, AllWISE, or {\em Spitzer} missions, which are not included in SIMBAD.

\subsection{NED versus SIMBAD: classifications}

\begin{figure}
    \centering
    \includegraphics[width=\columnwidth]{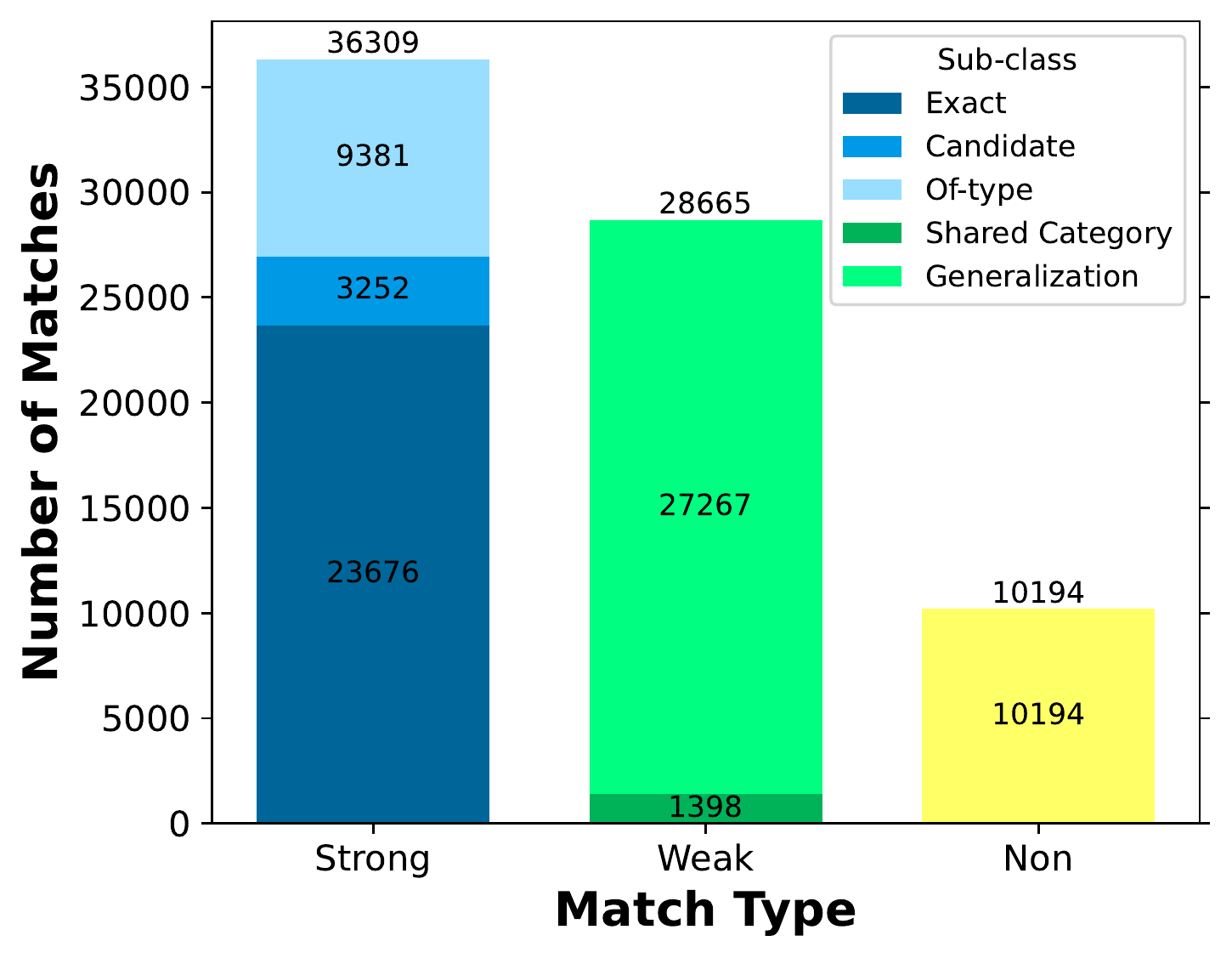}
    \caption{Distribution of match sub-classes for nearby galaxies. 
    Sub-classes of matches include exact matches (e.g. \hii region matched to \hii region), candidate matches (e.g. supernova remnant matched to supernova remnant candidate), of-type matches (e.g. galaxy matched to AGN), shared-category matches (e.g. cluster of stars matched to possible globular cluster), generalization matches (e.g. infrared source matched to star), and non-matches (e.g. star matched to galaxy).}
    \label{fig:Match_Type_Distribution}
\end{figure}

\begin{figure}
    \centering
    \includegraphics[width=\columnwidth]{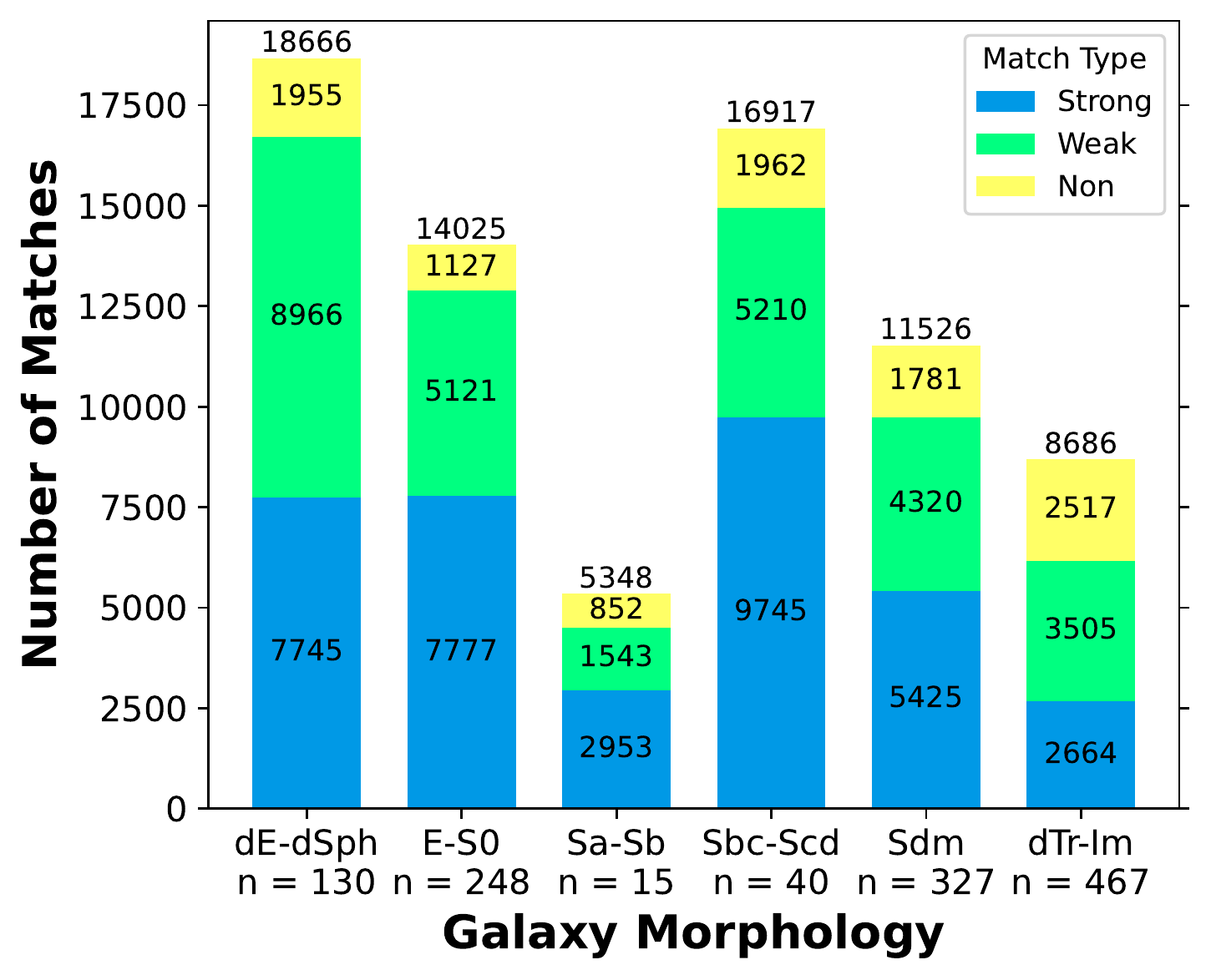}
    \caption{Classification comparison for matches between NED and SIMBAD objects for nearby galaxies, as a function of galaxy morphology characterized by de Vaucouleurs class.
    In order from left to right, the galaxy morphology labels comprise Hubble stage T values as follows: T = -3 with dwarf classifications dE or dSph; T = -3, -2, -1, 0; T = 1, 2, 3; T = 4, 5, 6; T = 7, 8, 9; T = -3 with dwarf classification dTr, 10, 11. 
    The $n$ value below each label represents the number of nearby galaxies in the bin.}
    \label{fig:Match_Type_GalMorph}
\end{figure}

Comparing the classifications for the matched objects, when taking into account both strong and weak match categories, NED and SIMBAD
were found to have an average classification overlap {\changed of 88\%}. 
When averaged across all galaxies, \changed{$\sim 66$\% of the compared NED and SIMBAD objects were classified as strong matches, $\sim 22$\%  were classified as weak matches, and $\sim 12$\% were classified as non-matches.
Of the strong matches, $\sim 65$\% were sub-classified as exact matches, $\sim 26$\%  were sub-classified as of-type matches, and  $\sim 9$\% were sub-classified as candidate matches.
Of the weak matches, $\sim 95$\% were sub-classified as generalization matches, and $\sim 5$\% were sub-classified as shared-category matches.}
(For a description of each match or sub-match type, refer to \autoref{sec:compare_class}.)
\autoref{fig:Match_Type_Distribution} shows the distribution of match classifications between NED and SIMBAD objects, providing a graphic representation of the above statistics. 
Strong matches make up the greater part of all match types and exact matches are found to be the predominant match sub-class, verifying our expectation that within the LVG, the majority of object classifications between NED and SIMBAD are equivalent.

\autoref{fig:Match_Type_GalMorph} examines the classification comparison of matches between NED and SIMBAD objects as a function of galaxy type characterized by de Vaucouleurs class. 
While the dwarf galaxies (dE--dSph) comprise the majority of NED/SIMBAD comparisons, the late type spiral galaxies (Sbc--Scd) have the most strong matches of any morphological category despite a significantly smaller number of source galaxies. 
Additionally, while the irregular galaxy category (dTm-Im) contains the largest number of source galaxies, it presents the second smallest number of NED/SIMBAD comparisons.
These conclusions are consistent with the findings above that that NED and SIMBAD contain more objects per galaxy for larger galaxies. 
We can infer that those individual objects are often better-studied in the larger galaxies, and in the most nearby dwarf galaxies, leading to more consistent classifications between NED and SIMBAD.
We also investigated the variation of match classifications with other galaxy properties including major linear diameter, distance, mass within the Holmberg radius, and $K_{\rm S}$-band luminosity and found no particularly noteworthy trends not covered by the above.

\subsection{Case studies}

Within the LVG sample, we selected two galaxies to further analyze through case studies.
Beginning with the 139 galaxies containing at least 1000 NED objects, we further selected the galaxy with the 
highest number of NED-SIMBAD matches that also had a classification \changed{match fraction of $\sim 99$\%} (UGCA~086) and the galaxy with the lowest classification match fraction (NGC~2903).
\autoref{tab:CS_table} summarizes the matching statistics for each of the case study galaxies. In both cases, the number of NED objects classified is an order of magnitude greater than the number of SIMBAD objects classified.

\begin{figure}
    \centering
    \includegraphics[width=\columnwidth]{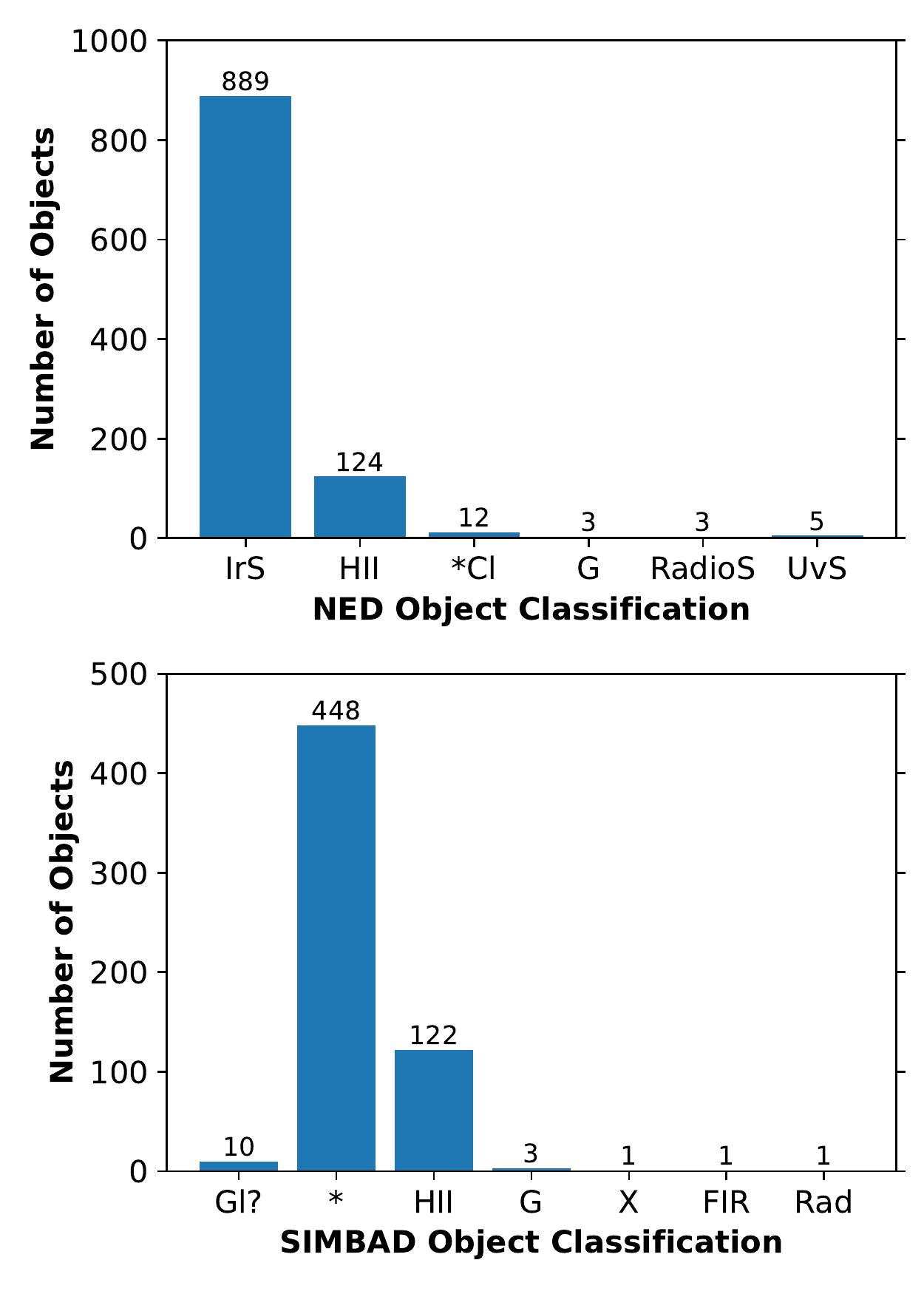}
    \caption{Comparison of the distribution of object classifications between NED and SIMBAD in the galaxy UGCA~086.}
    \label{fig:CS1_Object_Distribution}
\end{figure}

\begin{figure}
    \centering
    \includegraphics[width=\columnwidth]{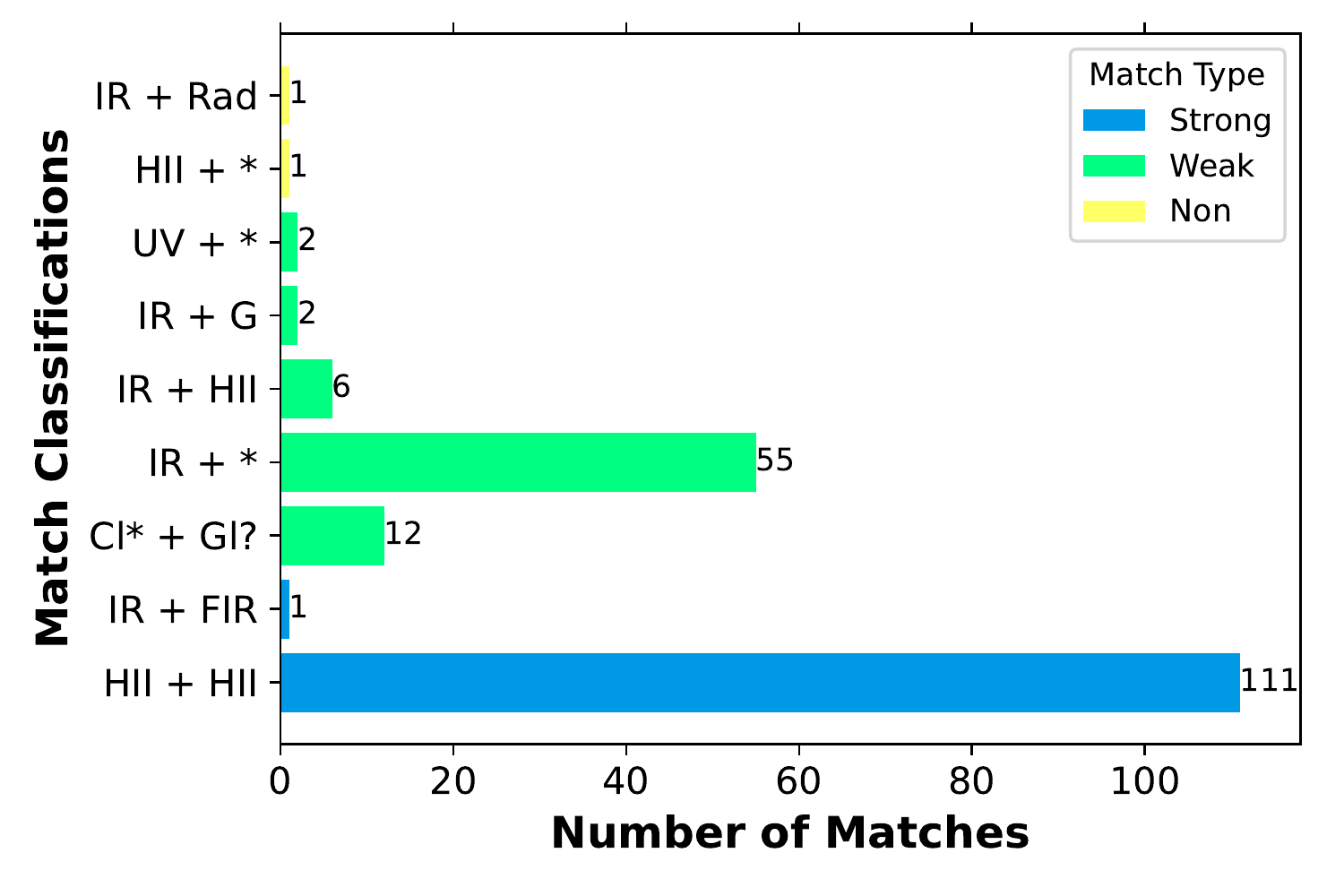}
    \caption{Distribution of match classifications in the galaxy UGCA~086. The bar colour here represents the overall match type of a given classification.}
    \label{fig:CS1_Match_Distribution}
\end{figure}

UGCA~086 is an irregular galaxy at a distance of 3~Mpc, located in the IC~342 group \citep{kara20}; 
it has substantial Galactic foreground extinction $A_B = 4.06$~mag and $L_K = 10^{9.1}$~L$_{\odot}$ \citep{kara13}.
All 98 of the NED and SIMBAD positionally-matched objects in this galaxy had matched classifications.
As \autoref{fig:CS1_Object_Distribution} shows, the positionally-matched objects are a small fraction of both the NED and SIMBAD contents for this galaxy: most of the 889 objects classified as `infrared source' by NED and most of the \changed{448} objects classified as `star' by SIMBAD are not positionally matched to objects in the other catalogue.
Figure \autoref{fig:CS1_Match_Distribution} shows the majority of object matches within the galaxy are strong matches \citep[\hii regions identified by][and included in both databases]{1996AJ....112.1096M}, while the few weak matches are largely forms of generalized matches such as `infrared source' (IR) to `galaxy' (G) or `ultraviolet Source' (UV) to `star' (*).

\begin{figure}
    \centering
    \includegraphics[width=\columnwidth]{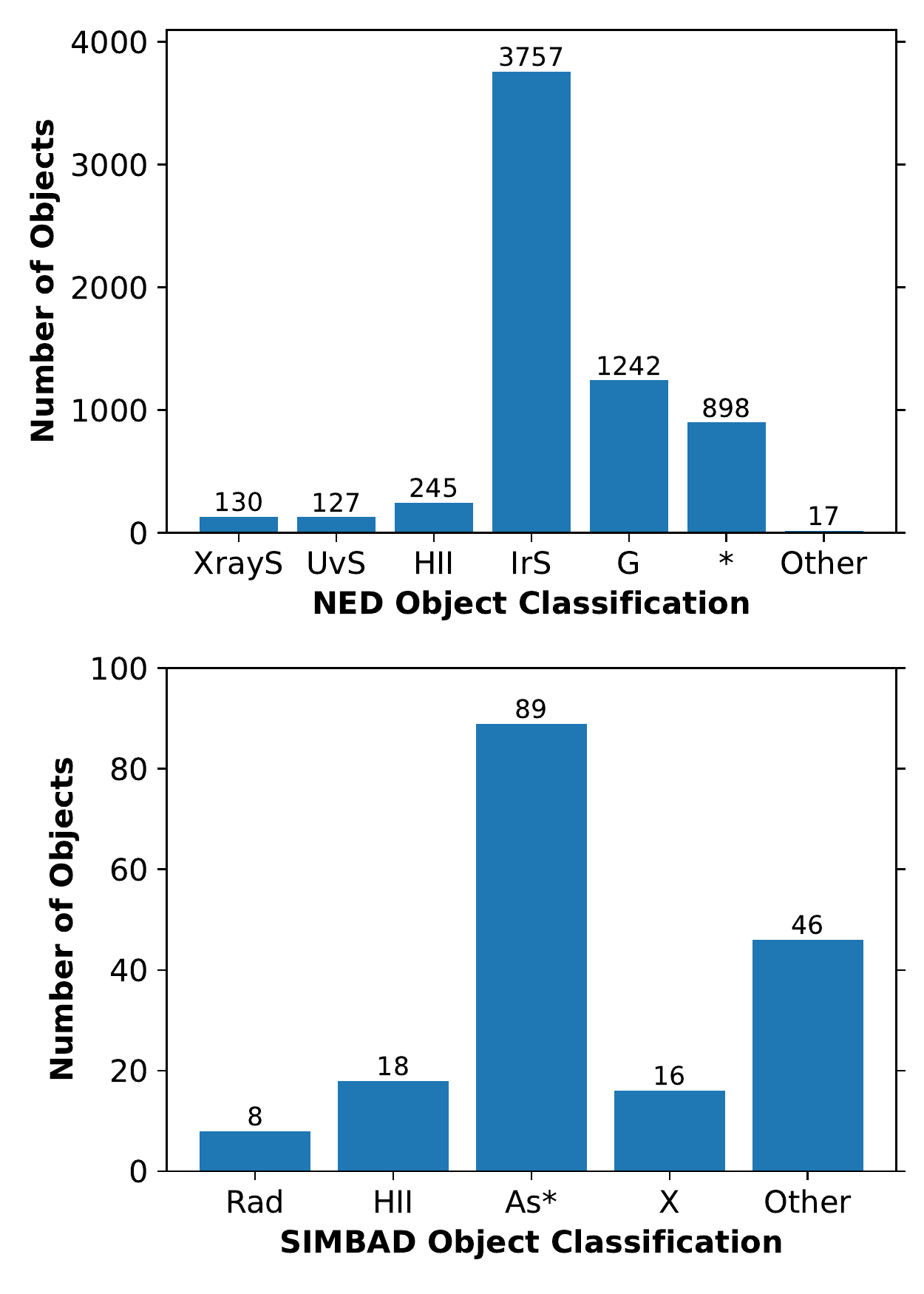}
    \caption{Comparison of the distribution of object classifications between NED and SIMBAD in the galaxy NGC~2903.}
    \label{fig:CS2_Object_Distribution}
\end{figure}

NGC~2903 is a large, relatively isolated spiral galaxy \citep{irwin2009} at a distance of 8.89~Mpc,  
with a luminosity $L_K = 10^{10.9}$~L$_{\odot}$ \citep{kara13}.
This galaxy has a notably low total match fraction in comparison to other galaxies within the LVG: \changed{only about 35\% of} the NED and SIMBAD objects compared within NGC~2903 were determined to have the same classification.
\autoref{fig:CS2_Object_Distribution} illustrates the comparison of object classifications between NED and SIMBAD for NGC~2903. As with UGCA~086, most of the NED `infrared sources' are not matched to SIMBAD sources; NED also contains {\changed about 150} times as many stars as does SIMBAD for this galaxy.

\begin{figure}
    \centering
    \includegraphics[width=\columnwidth]{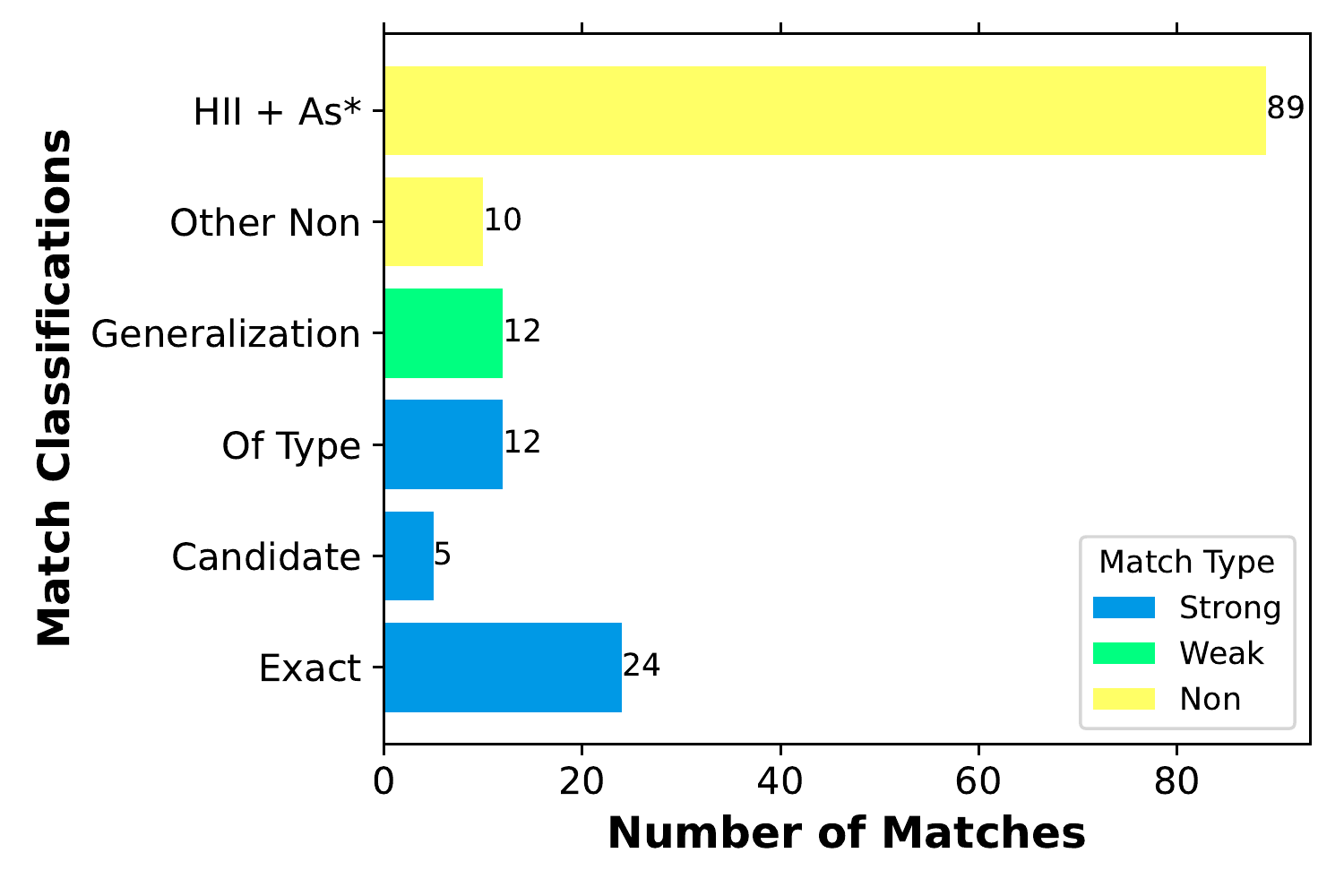}
    \caption{Distribution of match classifications in the galaxy NGC~2903. The bar colour here represents the overall match type of a given classification.}
    \label{fig:CS2_Match_Distribution}
\end{figure}

\autoref{fig:CS2_Match_Distribution} shows that almost all of the classification non-matches between these two galaxies come as a result of a single conflicting object classification: \hii region (HII) as classified by NED matched to stellar association (As*) as classified by SIMBAD.
Examination of the original identifications for these objects shows that in fact the original source for both databases is the work by \citet{2009MNRAS.396.2295H}. 
That work is a study of the ionizing stellar populations responsible for \hii region gaseous emission in the circumnuclear region of NGC~2903.
The individual positions catalogued by NED and SIMBAD are labelled as `regions' rather than specifically \hii regions or stellar associations. 
This ambiguity in the source material proves to be an example  of an instance where neither the NED or SIMBAD is incorrect in their object classification -- they simply made separate decisions which resulted in the occurrence of a classification non-match when in fact the same objects were being compared in both databases.

\begin{table*}
	\centering
	\caption{NED and SIMBAD matching in case study galaxies}
	\label{tab:CS_table}
	\begin{tabular}{llllllll} 
	\hline
	Name & $N_{\rm NED}$ & $N_{\rm SIMBAD}$ & $N_{\rm compared}$ & $N_{\rm match}$ & $N_{\rm strong}$ & $N_{\rm weak}$ & $N_{\rm non}$ \\
	\hline
	UGCA~086 & 1036 & 586 & 191 & 189 & 112 & 77 & 2 \\
	NGC~2903 & 6416 & 177 & 152 & 53 & 41 & 12 & 99 \\
	\hline
	\end{tabular}
\end{table*}

\section{Discussion}

This work has the following limitations.
Weak match types, such as generalization and same-category matches, are by their nature ambiguous.
The non-match sub type in its definition does not necessarily guarantee object classifications are incompatible, but rather that no match type is defined for the given object pair. Thus, it could be possible for two analogous objects to be classified as a non-match due only to an ambiguity in the two databases' overall classification scheme or in our determination of the correspondence between them.
Some of the objects categorized as non-matches could in fact be different objects that happen to lie within \changed{5~arcsec}, as opposed to the same object classified differently in the two databases.

Furthermore, as exemplified by our case studies, it is possible that in some instances, ambiguity in the source material used by NED and SIMBAD in their object classification could allow for certain matches to be incorrectly classified. This was illustrated in our study of NGC~2903 where 89 object matches (\hii to As*) were classified as non-matches due to different interpretations of the same source material. Although the primary source of this result was minor differences in object classification between NED and SIMBAD, it is important to recall that real astrophysical objects do not fit neatly into perfect boxes.

\section{Conclusions}

We find that, at the present time, NED contains a more comprehensive listing of objects in the vicinity of nearby galaxies. 
When totalling and averaging the results across the galaxy sample, we find about 11 times and 28 times the number of NED objects for every SIMBAD object classified.
Given the order-of-magnitude larger number of NED objects, the overlap between NED and SIMBAD is substantial, with \changed{about two thirds} of SIMBAD objects having corresponding entries in NED.

Of the positional matches, we find that \changed{88\%} of the NED/SIMBAD classifications agree, confirming that the vast majority of objects have similar classifications in both databases. More specifically,  \changed{$\sim 66$\% of the compared NED and SIMBAD objects were classified as strong matches whereas only $\sim 22$\% were classified as weak matches and $\sim 12$\% as non-matches.} 
These quantities do not depend strongly on the properties of the parent galaxies such as major angular diameter at the Holmberg isophote, indicative mass within the Holmberg radius, or $K_{\rm S}$-band luminosity.
The compatibility between NED and SIMBAD is reassuring and not unexpected given the wide use and careful compilation of these databases.
For the specific case of nearby galaxies, we find that many sources --- especially those associated with a particular galaxy as opposed to an all-sky survey --- are contained in only one of NED or SIMBAD. 
We recommend that researchers seeking the most complete picture of an individual galaxy's contents are best served by combining information from both databases.

\section*{Acknowledgements}
We thank the anonymous referee for helpful comments that improved the presentation of this work.
We acknowledge helpful discussions with C. Dewsnap and A. Man.
PB acknowledges support from an NSERC Discovery Grant.
This research has made use of the NASA/IPAC Extragalactic Database (NED), which is operated by the Jet Propulsion Laboratory, California Institute of Technology, under contract with the National Aeronautics and Space Administration.
This research has made use of the SIMBAD database, operated at CDS, Strasbourg, France.

\section*{Data availability}

The data underlying this article are available in Zenodo, via {\url{https://doi.org/10.5281/zenodo.5965260}}. 
The datasets were derived from sources in the public domain: NED {\url{http://ned.ipac.caltech.edu/}}, SIMBAD {\url{https://simbad.u-strasbg.fr/simbad/}}, and the Local Volume Galaxy catalogue  \url{http://www.sao.ru/lv/lvgdb/}.

\bibliographystyle{mnras}
\bibliography{NED_SIMBAD_class} 

\appendix

\section{NED and SIMBAD classification correspondence}

\autoref{tab:ned_classes} lists the NED object classes and the SIMBAD analogues that produce exact matches.
An exclamation mark in NED classes indicates that the object is in the Milky Way; SIMBAD does not make this distinction. We refer the reader to the NED and SIMBAD documentation for extended explanations of all object abbreviations.

\changed{As this work was in the final stages of the refereeing process, the SIMBAD classifications were reorganised;}%
\footnote{\url{https://simbad.cds.unistra.fr/guide/otypes.htx}} 
\changed{our analysis does not reflect this re-organisation.
Changes to the SIMBAD hierarchy could potentially affect two of our match types (of-type and shared-category), representing 15\% and $<1$\% of all classifications comparisons, respectively.
We expect that only a fraction of these comparisons would change match type, so we do not expect a major change to our results with the new SIMBAD hierarchy. 
}

To more accurately represent the \changed{(previous)} SIMBAD classification hierarchy in our program, we made some minor changes to SIMBAD object classification numerical codes. 
``T Tau Star Candidate'' was changed from 10.12.02.03 (10.12.00.00 is the ``Possible Peculiar Star'' class; however there is no 10.12.02.00 subclass) to 10.12.01.03 (a subclass of ``Young Stellar Object Candidate'').
 ``T Tau-type Star'' was changed from 14.06.25.03 to 14.06.25.00 as there was no 14.06.25.00 parent class.
The object class ``Circumstellar Matter'' (numerical code 13.14.00.00), which appears to have been removed in a change to the hierarchy, was added back to the classification scheme to act as a parent class for objects ``Outflow Candidate'' (numerical code 13.14.01.00), ``Outflow'' (numerical code 13.14.15.00), and ``Herbig-Haro Object'' (numerical code 13.14.16.00).  	 

\begin{table}
	\centering
	\caption{NED object classes with associated SIMBAD analogue classes.} 

	\label{tab:ned_classes}
	\begin{tabular}{ll} 
	\hline
	NED Classification & SIMBAD Analogue \\
	\hline
	* & * \\
    ** & ** \\
    *Ass & As* \\
    *Cl & Cl* \\
    AbLS & None \\
    Blue* & BS* + s*b \\
    C* & C* \\
    EmLS & Em* + EmG \\
    EmObj & EmO \\
    exG* & * \\
    Flare* & Er* \\
    G & G \\
    GammaS & gam \\
    GClstr & ClG \\
    GGroup & GrG \\
    GPair & PaG \\
    GTrpl & GrG \\
    GLens & LeG \\
    HII & HII \\
    IrS & IR \\
    MCld & MoC \\
    Neb & Cld \\
    Nova & No* \\
    Other & ? + err \\
    PN & PN \\
    PofG & PoG \\
    Psr & Psr \\
    QGroup & None \\
    QSO & QSO \\
    QLens & LeQ \\
    RadioS & Rad \\
    Red* & RG* + s*r \\
    RfN & RNe \\
    SN & SN* \\
    SNR & SNR \\
    UvES & UV \\
    UvS & UV \\
    V* & V* \\
    VisS & None \\
    WD* & WD* \\
    WR* & WR* \\
    XrayS & X \\
    !* & * \\
    !** & ** \\
    !*Ass & As* \\
    !*Cl & Cl* \\
    !Blue* & BS* + s*b \\
    !C* & C* \\
    !EmObj & EmO \\
    !Flar* & Er* \\
    !HII & HII \\
    !MCld & MoC \\
    !Neb & GNe \\
    !Nova & No* \\
    !PN & PN \\
    !Psr & Psr \\
    !RfN & RNe \\
    !Red* & RG* + s*r \\
    !SN & SN* \\
    !SNR & SNR \\
    !V* & V* \\
    !WD* & WD* \\
    !WR* & WR* \\
	\hline
	\end{tabular}
	
\end{table}

\bsp	
\label{lastpage}
\end{document}